%% file: paper.tex
\documentclass[11pt]{article}

\usepackage[final]{acl}

\usepackage{times}
\usepackage{latexsym}
\usepackage[T1]{fontenc}
\usepackage[utf8]{inputenc}
\usepackage{microtype}
\usepackage{inconsolata}
\usepackage{graphicx}
\usepackage{booktabs}
\usepackage{multirow}
\usepackage{array}
\usepackage{listings}
\usepackage{xcolor}
\usepackage{amsmath}
\usepackage{amssymb}
\usepackage{dsfont}
\usepackage{xspace}
\usepackage[section]{placeins}
\usepackage{float}
\usepackage{capt-of}
\usepackage{stfloats}
\usepackage{needspace}

\newcommand{\bench}{\textsc{ExecRetrieval}\xspace}
\newcommand{\nqueries}{939\xspace}
\newcommand{\ncorpus}{4{,}694\xspace}
\newcommand{\ndistractors}{3{,}755\xspace}
\newcommand{\nmodels}{23\xspace}
\newcommand{\ntotalmodels}{24\xspace}

\title{\bench: Measuring the Functional-Correctness Gap in Code-Embedding Retrieval}

\author{Aaryan Kapoor \\
  Kennesaw State University \\
  \texttt{aaryankapoor2006@gmail.com} \And
  Md Abdullah Al Hafiz Khan \\
  Kennesaw State University \\
  \texttt{mkhan74@kennesaw.edu}}

\begin{document}
\maketitle

\begin{abstract}
Embedding-based code retrieval is a core component of coding agents and retrieval-augmented code generation, where retrieving \emph{correct} code matters more than retrieving lexically similar code. Existing code-retrieval benchmarks do not plant controlled, execution-verified single-edit variants of each query's canonical implementation in the search pool, leaving the question of whether embeddings can functionally discriminate correct from near-clone-but-incorrect code unanswered in a retrieval setting. Resolving this requires a benchmark whose search pool itself contains the relevant counterfactuals --- execution-verified buggy variants near-identical to each canonical --- so that a retriever's rank ordering can be directly tested for functional discrimination rather than topical or identity overlap. We introduce \bench, \nqueries\ Python tasks each paired with one execution-verified canonical implementation and up to four execution-verified buggy distractors, each generated by a mechanical mutation making a single targeted edit, and evaluate \nmodels\ dense embedding configurations plus BM25 under provider-native invocation with paired McNemar tests and query-level bootstrap intervals. With near-clone counterfactuals in the pool, the top hosted system reaches \texttt{exec@10}$=$1.00 but only \texttt{exec@1}$=$0.331; rank-1 misses are paired buggy variants 91.5--99.4\% of the time across the four leading systems, and the canonical scores below at least one of its four paired distractors in 67--78\% of queries on the leading systems. The full dataset, execution oracle, embedding matrices, environment snapshot, and pairwise statistical tests are released at the URL in Appendix~\ref{app:artifacts}.
\end{abstract}

\section{Introduction}
\label{sec:intro}

\begin{figure*}[!t]
\centering
\includegraphics[width=\textwidth]{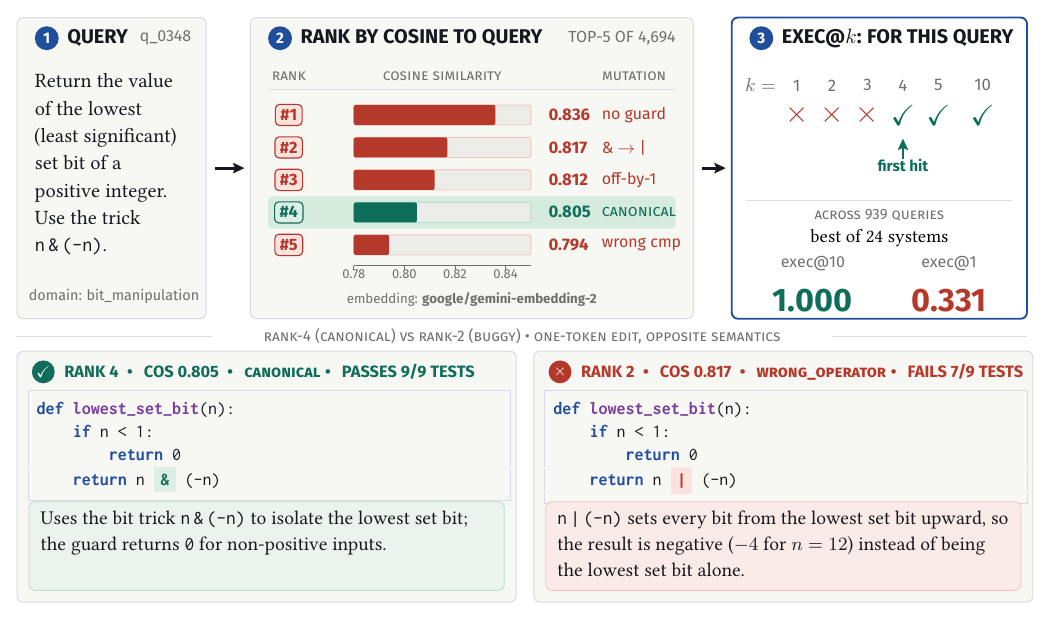}
\caption{\bench\ in one figure (query \texttt{q\_0348}, \texttt{lowest\_set\_bit}). The five paired snippets occupy the top-5 retrieval slots under Gemini Embedding~2 (best of \ntotalmodels\ systems); the canonical lands at rank 4 because three buggy variants embed closer to the query. Across all \nqueries\ queries, the best embedder reaches \texttt{exec@10}$=$1.000 yet \texttt{exec@1}$=$0.331; rank-1 misses are paired buggy variants 91.5--99.4\% of the time across the four leading systems.}
\label{fig:overview}
\end{figure*}

Embedding-based retrieval is becoming central to practical software engineering: coding assistants retrieve snippets from local repositories and the wider open-source corpus to ground their suggestions \citep{zhang2023repocoder,wang2024coderagbench}; IDEs use embeddings to locate prior occurrences of a function; and retrieval-augmented generation pipelines for code \citep{lewis2020rag,zhang2023repocoder} rely on embeddings to surface relevant context. In all of these settings, the embedding model serves as the candidate-recall stage of a longer pipeline: correctness is ultimately checked downstream by reranking, execution, tests, or agentic post-processing, but the first-stage ranking determines which candidates those later stages ever see, and in what order. What has not been measured is how much of the correctness-discrimination work this first stage passes on to the rest of the system when a near-identical buggy variant of the correct implementation sits in the search pool.

There is a measurement gap. The axis we care about --- whether the retriever surfaces an implementation that actually passes a test suite when a near-identical buggy variant is one cosine-similarity step away --- is not isolated as a controlled measurement anywhere. Public code-retrieval benchmarks score against canonical identity and topical similarity on found code: CodeSearchNet \citep{husain2019codesearchnet}, CodeXGLUE \citep{lu2021codexglue}, CoSQA \citep{huang2021cosqa}, CoIR \citep{li2025coir}, and CodeRAG-Bench \citep{wang2024coderagbench} all operate above this axis. None of them pair every canonical with controlled single-edit buggy variants in the search pool, so a perfect topical retriever and a perfect functional retriever score identically on them, even though the two notions disagree on the cases that matter for downstream code use.

This paper isolates the functional-correctness gap. Figure~\ref{fig:overview} shows a concrete instance: for the query \texttt{lowest\_set\_bit}, the strongest hosted embedder ranks three execution-failing buggy variants above the canonical implementation, which passes all of its tests yet lands only at rank~4. Concretely, we ask: \emph{when each natural-language query in a code-retrieval benchmark has both a canonical implementation and several near-identical buggy variants in the search pool, can modern embedding models still place a passing implementation at the top of the ranking?} Every magnitude we report is conditional on that premise --- a near-clone candidate present in the pool --- not an estimate of how often deployed corpora pose this choice; Section~\ref{sec:ablation} measures how the gap scales with density, and Limitations states the scope.

Closing this gap requires a benchmark with four properties: (i) for each query, a search pool containing both a known-correct implementation and a small set of near-identical buggy variants; (ii) correctness verified by execution against a deterministic test suite, not by human labelling; (iii) coverage across enough algorithmic domains and mutation archetypes that failure modes cannot be attributed to a single domain or bug pattern; and (iv) per-query paired statistics, so that retrievers are compared on the same inputs rather than on aggregate scores. We refer to a benchmark satisfying these as \emph{functionally grounded} code retrieval; \bench\ is one instantiation.

\needspace{3\baselineskip}
\paragraph{Contributions.}
\begin{enumerate}
\item \textbf{Dataset.} We construct \bench, a Python code-retrieval benchmark of \nqueries\ tasks across ten algorithmic domains. Each task ships with one canonical implementation, up to four mechanically mutated buggy distractors (one query carries 3, the rest 4), and a 7--10-test execution oracle. All canonicals pass all of their own tests; all distractors fail at least one. Total corpus: \ncorpus\ snippets.
\item \textbf{Construction pipeline.} We describe a high-yield reasoning-LLM pipeline (GPT-5.4 with high reasoning effort) with a 91\% first-attempt validation rate (91 of 100 validation-batch entries). The pipeline includes a five-stage validation gate (schema, AST semantics, canonical execution, distractor execution, corpus integrity) and an isolated-subprocess execution runner.
\item \textbf{Provider-native evaluation.} We evaluate \nmodels\ dense embedding configurations plus BM25 under each model's documented best-fair-shot invocation (task types, query and passage prefixes, dtype, normalization, batch size, and similarity metric, all sourced from primary provider documentation). We report \texttt{exec@k} and \texttt{execution\_precision@k} alongside canonical-ID nDCG (Normalized Discounted Cumulative Gain), with paired McNemar tests and query-level bootstrap intervals.
\item \textbf{Empirical findings.} Top-$k$ embedding retrieval is surprisingly strong (top hosted system reaches \texttt{exec@10}=1.00 across all \nqueries\ queries), but rank-1 retrieval is weak: 33.1\% at best. The same canonical implementation scores lower in query-cosine than at least one of its four paired buggy variants in 66.8\% of queries on Gemini Embedding~2 and 78.4\% on Qwen3-Embedding-8B. When rank-1 is wrong, it is almost always a paired mutation (91.5--99.4\% across the four leading systems). The deception rate is broadly distributed across the six mechanical mutation types: 44.3\% overall, spanning a 39.3\%--48.0\% band, with \texttt{remove\_edge\_case\_check} the least deceptive and \texttt{wrong\_comparison} the most.
\end{enumerate}

\section{Related Work}
\label{sec:related}

\paragraph{Code-retrieval benchmarks.} The shared gap relevant to this work is that no prior code-retrieval benchmark pairs every canonical with controlled, execution-verified single-edit variants in the search pool, so the effect of a specific near-clone on rank ordering cannot be isolated from topical or identity overlap. This applies whether the benchmark is human-annotated (CodeSearchNet's challenge split \citep{husain2019codesearchnet} with $\sim$4{,}000 expert relevance labels over 99 queries; CoSQA \citep{huang2021cosqa} with 20{,}604 human-annotated query-code labels), an aggregation suite (CodeXGLUE \citep{lu2021codexglue} bundles 14 datasets across 10 tasks; CoIR \citep{li2025coir} aggregates 10 datasets across 8 retrieval tasks), or downstream-of-retrieval (CodeRAG-Bench \citep{wang2024coderagbench} measures generation quality after retrieval). CoQuIR \citep{geng2025coquir} is the closest prior work on the quality axis, with 42{,}725 queries over 134{,}907 snippets annotated for correctness, efficiency, security, and maintainability; its correctness labels come from online-judge verdicts and Defects4J bug/fix pairs over naturally occurring code, whereas \bench\ plants controlled single-edit mutants of each query's own canonical, so every wrong retrieval is one attributable, test-checkable edit. xCodeEval \citep{khan2024xcodeeval} scores NL-to-code retrieval by execution, with wrong-answer submissions as negatives, over naturally occurring solutions rather than controlled edits of each query's canonical. Closest in spirit on the correctness axis, \citet{li2025functional} also study syntactically similar but functionally divergent code: an LLM-driven synthesis framework evolves Type-IV variants (similar syntax, different function) validated against generated test suites, and fine-tuning embedding models on these pairs improves pairwise functional-consistency classification and code-to-code retrieval over a standard pool \citep[xCodeEval;][]{khan2024xcodeeval}. The distinction is construction-level: their execution-validated variants are evaluated pairwise and serve as fine-tuning data, and retrieval is measured code-to-code, whereas \bench\ plants each execution-verified counterfactual inside the natural-language retrieval pool itself, so what is tested is a retriever's rank ordering over query-to-code search, with every failure attributable to a single targeted edit.

\paragraph{Execution-based evaluation.} On the generation side, HumanEval and \texttt{pass@k} \citep{chen2021codex} established the pattern of testing whether sampled code passes a held-out test suite; HumanEval+/EvalPlus \citep{liu2023evalplus} showed strengthening tests materially reduces reported \texttt{pass@k}. MBPP \citep{austin2021mbpp}, DS-1000 \citep{lai2023ds1000}, CRUXEval \citep{gu2024cruxeval}, LiveCodeBench \citep{jain2024livecodebench}, and SWE-bench \citep{jimenez2024swebench} extend this idea up to whole-repository issue resolution. We adopt the same execution-grounded discipline for the \emph{retrieval} step that precedes generation.

\paragraph{Embedding models.} We treat the current generation of code-aware and general-purpose embedding models as black-box services. Code-trained models include CodeBERT \citep{feng2020codebert}, GraphCodeBERT \citep{guo2021graphcodebert}, CodeT5+ \citep{wang2023codet5plus}, ContraCode \citep{jain2021contracode}, and CodeRetriever \citep{li2022coderetriever}. General-purpose dense retrievers we evaluate trace to Sentence-BERT \citep{reimers2019sbert}, DPR \citep{karpukhin2020dpr}, and the OpenAI \citep{neelakantan2022openai}, E5 \citep{wang2022e5}, GTE \citep{li2023gte}, BGE \citep{xiao2024bge,chen2024m3}, Qwen3 Embedding \citep{zhang2025qwen3}, and Gemini Embedding \citep{lee2025gemini} families. MTEB \citep{muennighoff2022mteb} established large-scale embedding evaluation; domain-specific MTEB-style work shows general-domain rankings transfer poorly to specialized retrieval \citep{tang2025finmteb}.

\section{The \bench\ Dataset}
\label{sec:dataset}

\subsection{Task and structure}
Informally, an \bench\ instance gives a model a natural-language description of a function and asks it to rank one execution-verified canonical implementation above four near-identical buggy variants of that canonical, where ``near-identical'' means the variants differ from the canonical by a single targeted edit. Formally, an instance is a tuple
\[
(\, q,\; f,\; c,\; T,\; D_1,\dots,D_n\, ),
\]
where $q$ is a natural-language description of a function, $f$ is the target function name, $c$ is a canonical Python implementation, $T = (t_1,\dots,t_m)$ with $7 \le m \le 10$ is an executable test suite of \texttt{assert} statements, and $D_1,\dots,D_n$ with $n\in\{3,4\}$ are single-mutation buggy distractors of $c$. We require $c$ to pass every $t_i$ and each $D_j$ to fail at least one $t_i$. The retrieval corpus is the union of all canonicals and distractors across the benchmark: \ncorpus\ snippets ($\nqueries$ canonicals plus \ndistractors\ distractors; 938 queries contribute 4 distractors each and 1 query contributes 3, as one pilot-era distractor was retired in the validation audit).

\begin{table}[!htb]
\centering
\small
\begin{tabular}{lr lr}
\toprule
Domain & \# queries & Domain & \# queries\\
\midrule
bit\_manip. & 97 & math\_num. & 90\\
collections & 98 & sort\_search & 94\\
data\_xform & 91 & state\_mach. & 84\\
date\_time & 99 & string\_proc. & 94\\
geometry & 94 & validation & 98\\
\midrule
\multicolumn{4}{r}{\textbf{Total: \nqueries}}\\
\bottomrule
\end{tabular}
\caption{Domain composition of \bench. Each entry contributes 1 canonical and up to 4 distractors to the retrieval corpus (one entry contributes 3).}
\label{tab:domain}
\end{table}

Tasks span ten algorithmic domains: bit-manipulation, collections, data-transformation, date-time, geometry, math/numerical, sorting/searching, state-machines, string-processing, and validation (Table~\ref{tab:domain}). Function names are globally unique across the registry, and each canonical defines exactly its target function plus optional locally scoped helpers.

\subsection{Generation pipeline}
\label{sec:generation}

Generation uses a two-phase, registry-driven pipeline. In the first phase, we prompt Claude Sonnet 4.6 \citep{anthropic_claude_sonnet} with high reasoning effort to produce 1{,}000 candidate \texttt{(function\_name, query, domain)} triples --- 100 per domain --- in a single completion, then manually deduplicate across domains in two passes (962 unique triples after the first pass, 954 after a second pass removed near-duplicates surfaced during pipeline iteration), shipping the 954-entry registry from which \nqueries\ validated entries were retained. Each registry entry specifies the target NL query and function name but contains no implementation; the LLM that later writes implementations cannot collude with itself on task choice. In the second phase, for each registry triple, we issue one API call per attempt to GPT-5.4 \citep{openai_gpt54} with \texttt{reasoning\_effort=high} through the OpenAI API, asking the model to return JSON with a canonical implementation, a 7--10-statement \texttt{assert}-only test suite, and four mechanical-mutation distractors, each making a single targeted edit. The locked prompt forbids \texttt{wrong\_semantics} (``write an alternative algorithm'') and constrains distractors to six bug types: \texttt{off\_by\_one}, \texttt{wrong\_operator}, \texttt{swap\_arguments}, \texttt{remove\_edge\_case\_check}, \texttt{wrong\_comparison}, \texttt{off\_by\_one\_boundary}. These correspond closely to classical mutation-testing operators --- constant replacement, relational/arithmetic operator replacement, and statement deletion \citep{jia2011mutation} --- a family shown to produce faults that couple to real ones for testing purposes \citep{just2014mutants}. Each generated distractor also carries a free-text \texttt{bug\_description} field explaining the mutation; this field is part of the released dataset and provides a per-distractor explanation that downstream users can audit. The system prompt and the key user-prompt clauses are reproduced verbatim, along with a stylized excerpt of one validated entry, in Appendix~\ref{app:prompts}; the complete prompts and per-entry records are in the released \texttt{queries.jsonl}/\texttt{corpus.jsonl}. Aggregate token counts and the cost of phase~2 are in Appendix~\ref{app:economics}; per-entry counts ship in \texttt{batch\_usage}.

The choice of GPT-5.4 with high reasoning effort is decisive. Earlier iterations using Claude Sonnet~4 \citep{anthropic_claude4} and Claude Sonnet~4.6 (Sonnet~4 for the original pilot, Sonnet~4.6 for the first scaled run) produced \emph{accidentally correct} distractors in 127 of 400 cases ($31.8\%$): the model often emitted a correct alternative implementation when asked to write incorrect code, with 40 of those 127 cases self-admitting ``actually correct'' in the \texttt{bug\_description} field. The locked prompt forbids \texttt{wrong\_semantics} largely because of this failure mode --- almost every Sonnet correctness-compulsion case was logged under that bug type, and removing it from the allowed set forces the model to issue a single targeted mutation rather than rewriting the algorithm. GPT-5.4 with high reasoning all but eliminated the failure (3 of 4{,}112 first-attempt distractors, 0.07\%, passed their tests); \textbf{935 of the 939} released entries (99.6\%) come from GPT-5.4, the rest pilot-era Sonnet leftovers that passed all gates. Two further prompt-level refinements lifted the first-attempt validation rate to 91\% (91 of 100 validation-batch entries; failures were regenerated): an \emph{assert-only} test format directive eliminated 28\% of semantic rejects, and a \emph{must produce wrong output, not crash} clause cut crash-type distractor rejects from 24\% to 3\%. Each released entry carries a \texttt{metadata} field with model identifier, endpoint, and generation timestamp; 926 of 939 also carry per-entry token counts in \texttt{batch\_usage}, with latency on real-time entries only.

\subsection{Validation oracle and integrity audit}
\label{sec:validation}

Generated entries pass five sequential gates and are retained only if all five clear: \emph{(1) schema} (required fields, exactly 4 distractors, 7--10 tests); \emph{(2) AST semantics} (canonical and every distractor define the target function name; almost every test is a single \texttt{Assert} that calls it, with 15 of the 8{,}499 released tests using a multi-statement form); \emph{(3) canonical execution} (canonical passes every test); \emph{(4) distractor execution} (each distractor fails at least one test, 938/939 quads do not all fail on identical test subsets, no distractor is a pure syntax error, and 3{,}750/\ndistractors\ distractors produce at least one bare assertion failure so the dominant rejection mode is \emph{wrong output} rather than crashing on every input); and \emph{(5) corpus integrity} (every query's \texttt{correct\_corpus\_ids} reference exists; no correct canonical is unreferenced).

The validator drives execution through a process-isolated runner. Each \texttt{(code, test\_suite)} pair runs in a single fresh Python subprocess (\texttt{-I}, minimal env) that iterates the suite with a fresh namespace per test, with a 5-second per-suite timeout. Per-test outcomes are categorized into \texttt{pass}, \texttt{FAIL}, \texttt{FAIL:<ExceptionType>}, \texttt{TIMEOUT}, and \texttt{ERROR:<ExceptionType>}. The runner injects a minimal \texttt{pytest.raises} shim so exception-raising tasks are evaluable without a full pytest dependency. Final state: \textbf{939/939} canonicals pass their own tests; \textbf{\ndistractors/\ndistractors} paired distractors fail at least one test; the execution cache contains \textbf{46{,}458} rows, one per distinct \texttt{(code, test\_suite)} pair encountered across validation, the cross-canonical integrity sweep below, and top-10 scoring for every retrieval system; each row stores a per-test outcome list. One pilot-era distractor (bug type \texttt{boundary\_error}, query \texttt{q\_0001}) was retired during a late audit because it crashed on every input with \texttt{NameError} (malformed code) rather than failing on output; \ndistractors\ paired distractors remain.

\begin{table}[!htb]
\centering
\small
\setlength{\tabcolsep}{4pt}
\begin{tabular}{lrr}
\toprule
Bug type (locked prompt) & count & \%\\
\midrule
\texttt{wrong\_operator} & 857 & 22.8\\
\texttt{off\_by\_one\_boundary} & 650 & 17.3\\
\texttt{swap\_arguments} & 650 & 17.3\\
\texttt{off\_by\_one} & 642 & 17.1\\
\texttt{wrong\_comparison} & 617 & 16.4\\
\texttt{remove\_edge\_case\_check} & 329 & 8.8\\
\midrule
Pre-lock legacy (4 types, combined) & 10 & 0.3\\
\midrule
\textbf{Total} & \textbf{\ndistractors} & \textbf{100.0}\\
\bottomrule
\end{tabular}
\caption{Distractor distribution by bug type in the released corpus. The six mechanical mutation types of the locked prompt account for 99.7\% of distractors; the remaining 10 entries are pilot-era legacy bug types that survived all validation gates --- \texttt{ignores\_constraint}, \texttt{missing\_edge\_case}, \texttt{wrong\_semantics} (3 each); \texttt{wrong\_algorithm} (1).}
\label{tab:bugtypes}
\end{table}

Because mechanically mutated distractors are similar to their canonical, test-suite ambiguity could corrupt the deception measurement. We ran two integrity sweeps: a \emph{distractor-sanity} sweep, in which every one of the \ndistractors\ paired distractors is executed against its own query's tests, and a \emph{cross-canonical} sweep, which exhaustively searches every corpus item's AST for a function definition whose name matches any \emph{other} query's target function and, when found, runs that item against the other query's tests. The first sweep finds 0 distractors that pass all of their own tests; the second sweep finds 0 module-level cross-entry name collisions in the released corpus, so no canonical or distractor can accidentally satisfy another query's tests. The 939 target function names are globally unique across the registry. Distractor bug-type composition is given in Table~\ref{tab:bugtypes}.

\section{Experimental Setup}
\label{sec:protocol}

\subsection{Metrics}
For a query $q$, let $\hat{r}_1,\dots,\hat{r}_{|C|}$ be the corpus retrieval ranking and let $\textsc{pass}(c, T_q) \in \{0,1\}$ denote whether snippet $c$ passes all of $q$'s tests. Define:
\begin{align*}
\texttt{exec@k}(q) &= \mathds{1}\!\left[\, \exists\, i \le k : \textsc{pass}(\hat{r}_i, T_q) = 1\, \right],\\
\texttt{execp@k}(q) &= \frac{1}{k}\sum_{i=1}^{k} \mathds{1}[\textsc{pass}(\hat{r}_i, T_q) = 1].
\end{align*}
We report \texttt{exec@k} (at least one passing in top-$k$), \texttt{execution\_precision@k} (\texttt{execp@k}; fraction of top-$k$ that pass), and canonical-ID nDCG \citep{jarvelin2002ndcg}. Aggregates are taken as the unweighted mean over the \nqueries\ queries. All three families are reported for $k \in \{1,3,5,10\}$.

\subsection{Confidence intervals and paired tests}
Confidence intervals on each metric are computed by bootstrap resampling \citep{efron1979bootstrap} of the \nqueries\ queries with 5{,}000 replicates and fixed seed. For pairwise comparisons between models on a binary \texttt{exec@k}, we report the exact McNemar test \citep{mcnemar1947,dietterich1998mcnemar} on the discordant-pair counts, alongside the paired-bootstrap difference and its query-level 95\% interval. For continuous metrics (\texttt{execution\_precision}, nDCG) we report only the paired bootstrap interval. All raw test outputs are released alongside the dataset.

\subsection{Models and provider-native invocation}
\label{sec:invocation}

We evaluate \nmodels\ dense embedding configurations across Google Gemini, Mistral, OpenAI, Qwen3, BGE, BGE-M3, E5, GTE, and Sentence-Transformers families, plus a BM25 baseline \citep{robertson2009bm25} (\texttt{k1=1.5}, \texttt{b=0.75}). Each model is invoked with its documented best-fair-shot setup, taken from the provider's primary API docs or model card. Gemini Embedding~001 uses task-type \texttt{CODE\_RETRIEVAL\_QUERY} on queries and \texttt{RETRIEVAL\_DOCUMENT} on corpus snippets, per Google's documented code-retrieval recipe \citep{google_gemini_embeddings_docs}; Gemini~2's API does not expose task types, so we adopt the textual instruction conventions from the same documentation (``task: code retrieval'' for queries, neutral framing for corpus items). Qwen3 models prepend \mbox{``Instruct: $\langle$task$\rangle$\textbackslash nQuery: ''} to queries per the official model cards; E5 prepends \mbox{``query: ''/``passage: ''}; BGE prepends the official retrieval prefix to queries; OpenAI, Mistral, GTE, and Sentence-Transformers models receive raw text \citep{openai_embeddings_docs,mistral_embeddings_docs}. All models use cosine similarity over $L_2$-normalized embeddings except \texttt{multi-qa-mpnet-base-dot-v1}, which uses unnormalized dot product per its card. Qwen3-8B uses BF16 because FP16 produced non-finite similarities on our A40 hardware; we add Qwen3-4B-BF16 as a dtype control row. BGE-M3 is run in dense-only mode; hybrid dense+sparse+ColBERT would be a separate model row. Operationally critical settings appear in Appendix~\ref{app:invocation}, Table~\ref{tab:invocation}; the machine-readable invocation table (model IDs, dimensions, dtypes, similarity, batch sizes, conditioning, provider doc URLs) is released as \texttt{results/invocation\_table.json}; SDK versions are in \texttt{pip\_freeze.txt}.

\subsection{Reproducibility}
The frozen evaluation consists of: the \ncorpus-snippet corpus and \nqueries-query files (deterministic IDs, SHA-256 hashed); 23 saved \texttt{.npz} embedding matrices (each passes shape, finite-value, and zero-norm checks across queries and corpus); a per-pair execution-result cache; a \texttt{pip freeze}; and a SHA-256 manifest over all the above. The validator and runner depend only on the Python standard library plus a minimal \texttt{pytest.raises} shim. Provider API calls use retry/backoff and fail fast on missing API keys. The full bundle is released at the URL listed in Appendix~\ref{app:artifacts}.

\section{Results}
\label{sec:results}

\begin{figure*}[!tb]
\centering
\includegraphics[width=0.92\textwidth]{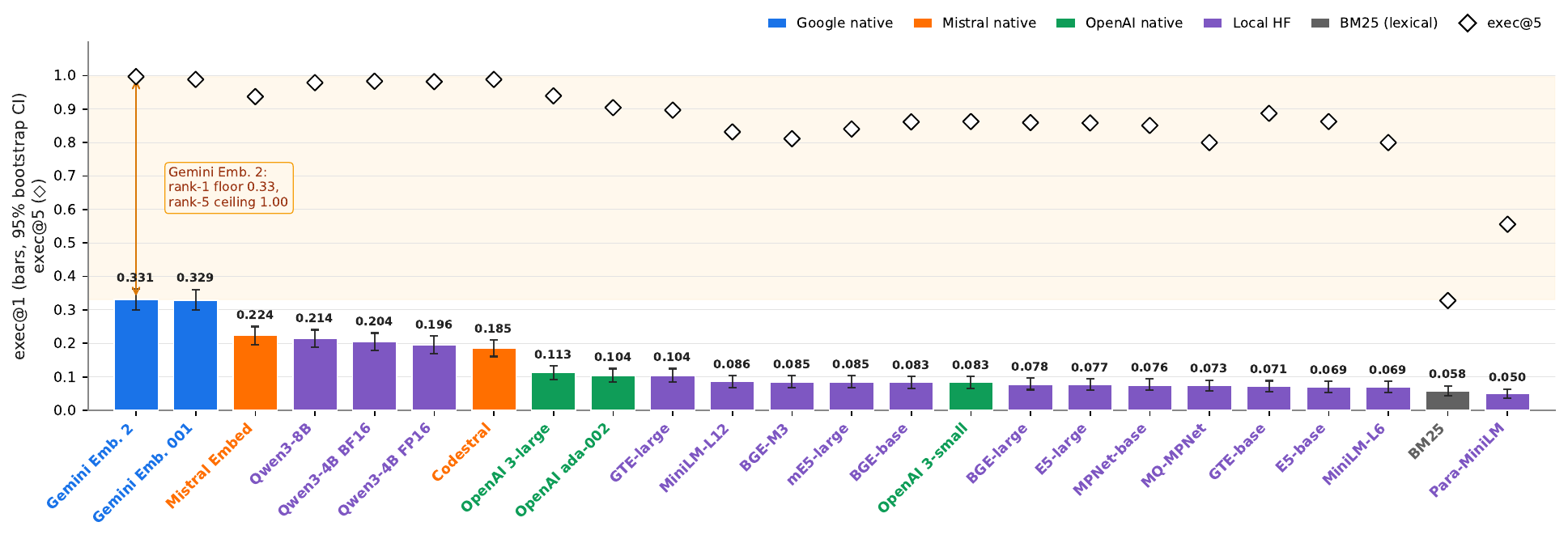}
\caption{\bench\ leaderboard. Bars are \texttt{exec@1} with 95\% bootstrap intervals; diamonds mark \texttt{exec@5}. Hosted Google Gemini Embedding models lead at rank~1; open-weight Qwen3 dominates at the top of the open-weight tier; and twelve of the thirteen embedding models below 10\% at rank~1 reach \texttt{exec@5} of 0.80--0.89.}
\label{fig:leaderboard}
\end{figure*}

\subsection{Top-$k$ saturates while rank-1 fails}

Figure~\ref{fig:leaderboard} shows the \nmodels-model dense leaderboard plus BM25, sorted by \texttt{exec@1}; the full per-model table with \texttt{exec@k}, \texttt{execution\_precision@k}, and nDCG@$k$ for $k\in\{1,3,5,10\}$ is reproduced in Appendix~\ref{app:full-leaderboard}. Three observations dominate.

\textbf{Rank-1 retrieval is hard.} The top \texttt{exec@1} system, Gemini Embedding 2, reaches only \texttt{exec@1}$=$0.331, with bootstrap 95\% CI $[0.299, 0.362]$. Gemini Embedding 001 reaches 0.329, and the two are statistically indistinguishable at \texttt{exec@1} (paired McNemar exact $p$=0.958; difference $+0.002$, 95\% CI $[-0.038, +0.042]$). Mistral Embed ranks third (0.224), followed by the open-weight Qwen3 family.

\textbf{Top-$k$ retrieval is easy.} \texttt{exec@10} is essentially saturated for hosted systems: Gemini Embedding 2 reaches \texttt{exec@10}$=$1.00 over all \nqueries\ queries, and Gemini Embedding 001 and Codestral Embed 2505 also achieve 1.00. Even mid-tier embedding models like BGE-base and E5-large-v2 reach \texttt{exec@10}~$\in$~[0.94, 0.95].

\textbf{Lexical retrieval is a poor proxy.} BM25 reaches \texttt{exec@1}$=$0.058 and \texttt{exec@10}$=$0.422. Lexical overlap occasionally identifies the canonical solution, but its top-10 success is below the worst dense embedding model (\texttt{paraphrase-MiniLM} at 0.671). BM25 is included as a calibration baseline, not as a competitive system.

\begin{figure}[!htbp]
\centering
\includegraphics[width=\columnwidth]{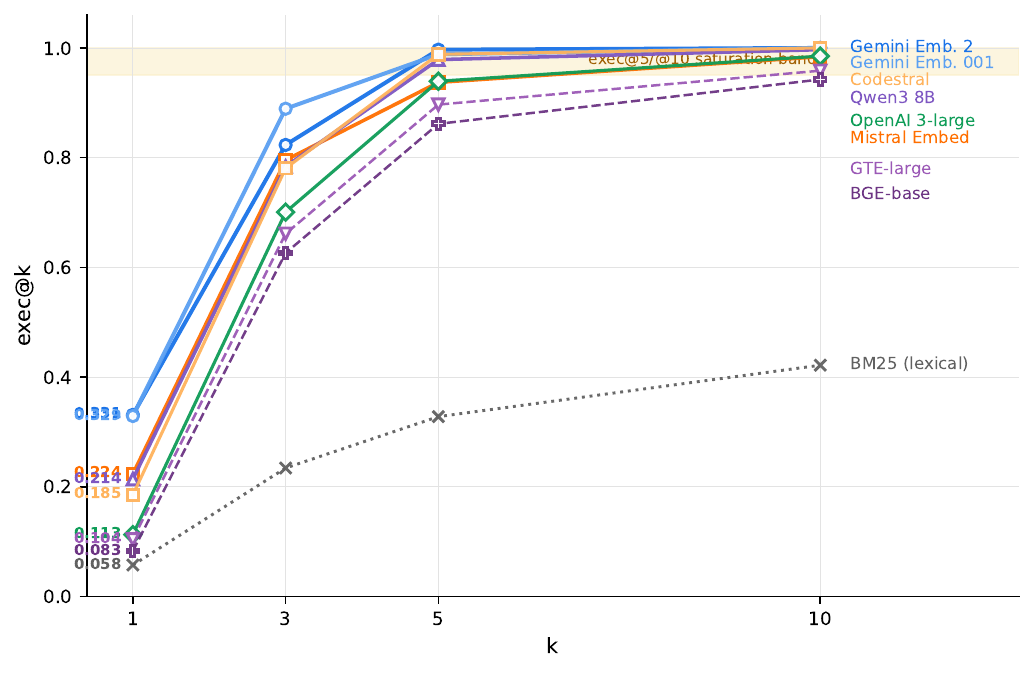}
\caption{\texttt{exec@k} curves for representative systems. Most strong systems saturate \texttt{exec@10} but plateau far below 1 at \texttt{exec@1}.}
\label{fig:execcurves}
\end{figure}

Figure~\ref{fig:execcurves} plots \texttt{exec@k} curves for representative systems and shows the structural pattern across the leaderboard: for hosted systems and Qwen3, the slope from $k$=1 to $k$=3 is steep, the curve has effectively saturated by $k$=5, and the rank-1 floor is much lower than top-$k$ headroom suggests.

\begin{figure*}[!tb]
\centering
\includegraphics[width=0.85\textwidth]{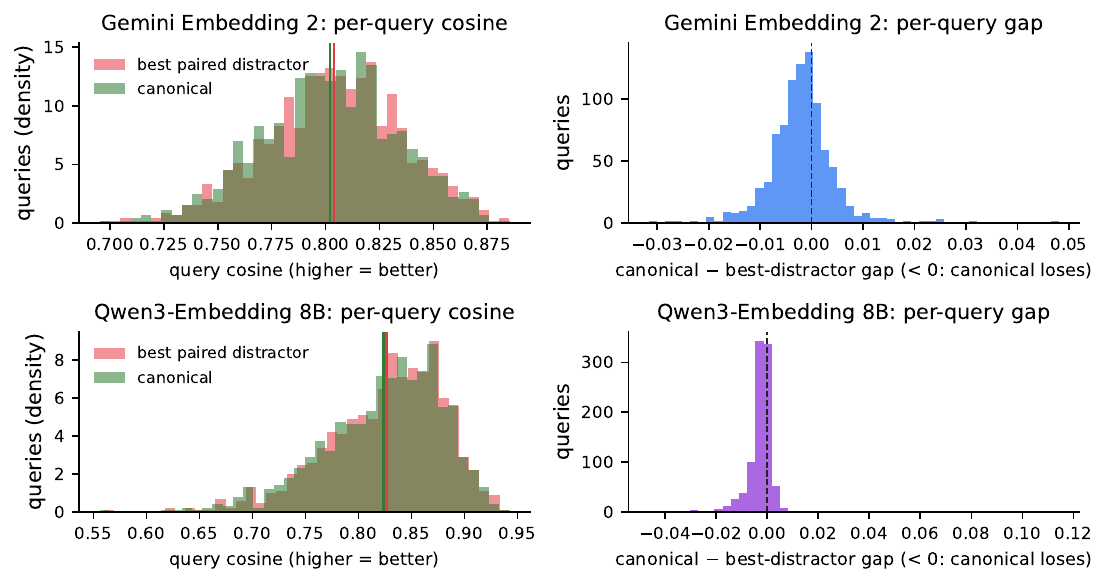}
\caption{Per-query cosine of the canonical (green) vs.\ highest-cosine paired distractor (red), for Gemini~2 (top) and Qwen3-8B (bottom); right column: per-query difference. The canonical lies below at least one paired distractor in 66.8\% and 78.4\% of queries.}
\label{fig:simgap}
\end{figure*}

\begin{figure}[!htbp]
\centering
\includegraphics[width=\columnwidth]{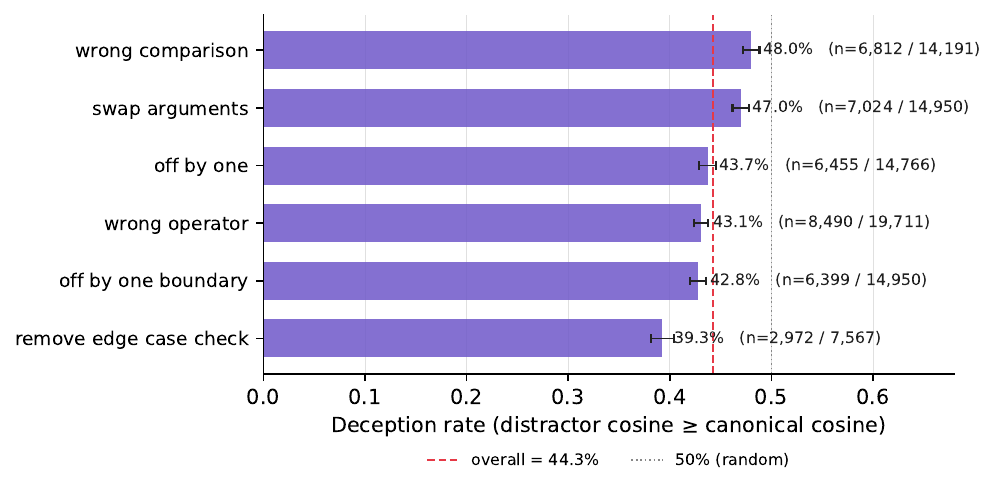}
\caption{Per-(query, paired distractor, model) deception rate by mechanical mutation type. A triple is ``deceived'' when the distractor's similarity to the query is at least as high as the canonical's. Aggregated over the \nmodels\ dense embedding configurations and the six mechanical mutation types defined in Section~\ref{sec:generation}.}
\label{fig:bugtype}
\end{figure}

\subsection{When the model misses, what does it retrieve?}
\label{sec:misses}

\begin{table}[!ht]
\centering
\small
\begin{tabular}{lrr}
\toprule
Model & miss@1 & paired dist.\ rank-1\\
\midrule
Gemini Embedding 2 & 628 & 624\,(99.4\%)\\
Gemini Embedding 001 & 630 & 618\,(98.1\%)\\
Mistral Embed & 729 & 667\,(91.5\%)\\
Qwen3-Embedding-8B & 738 & 718\,(97.3\%)\\
\bottomrule
\end{tabular}
\caption{When rank-1 is wrong, it is almost always a paired buggy variant. ``miss@1'' is the count of queries whose top-1 fails the tests; the last column is how many of those are paired single-mutation distractors of the same query's canonical.}
\label{tab:miss-anatomy}
\end{table}

To understand the structure of rank-1 failure, we ask, for each query where the rank-1 retrieved snippet does not pass the tests, \emph{which corpus item was placed at rank~1?} Two possibilities exist: it could be one of the four paired buggy variants of that query's canonical, or it could be any other snippet in the \ncorpus-snippet corpus (e.g., a canonical or distractor from a different entry). Because function names are globally unique and our integrity audit (Section~\ref{sec:validation}) shows no cross-canonical test-suite leakage, only the paired distractors are functional near-clones of the canonical; everything else is unrelated code.

Table~\ref{tab:miss-anatomy} shows the result. For all four leading models, \emph{between 91.5\% and 99.4\%} of rank-1 misses are paired buggy variants. Failure of embedding retrieval at rank~1 is therefore not an arbitrary lookup error; it is a specific, structural inability to separate a canonical from its single-mutation buggy near-clones.

\subsection{One near-clone is enough: pool-density ablation}
\label{sec:ablation}

\begin{table}[!htb]
\centering
\small
\setlength{\tabcolsep}{4pt}
\begin{tabular}{lccccc}
\toprule
 & \multicolumn{5}{c}{\texttt{exec@1} by \# paired distractors retained}\\
\cmidrule(lr){2-6}
Model & 0 & 1 & 2 & 3 & all\\
\midrule
Gemini Emb.\ 2 & 0.993 & 0.678 & 0.507 & 0.402 & 0.331\\
Gemini Emb.\ 001 & 0.984 & 0.718 & 0.541 & 0.418 & 0.329\\
Mistral Embed & 0.928 & 0.615 & 0.425 & 0.304 & 0.224\\
Qwen3-Emb.-8B & 0.976 & 0.613 & 0.409 & 0.290 & 0.214\\
\bottomrule
\end{tabular}
\caption{Pool-density ablation: expected \texttt{exec@1} when each query retains $d$ of its own paired distractors (exact expectation over all retained subsets; the rest of the \ncorpus-snippet corpus is unchanged). The ``all'' column is the shipped configuration and reproduces the published leaderboard exactly for every one of the \nmodels\ configurations. One near-clone drops the strongest system from 0.993 to 0.678.}
\label{tab:ablation}
\end{table}

Is the measured gap an artifact of the deliberately dense pool (up to four planted near-clones per query)? We ablate density directly: for each query we remove a subset of its own paired distractors, holding the rest of the corpus fixed, and compute the exact expected \texttt{exec@1} over all retained subsets of size $d \in \{0,\dots,4\}$ (the 3-distractor query contributes its shipped configuration at $d{=}4$). Ranking follows the scoring engine's realized sort order, so the full-pool column reproduces the published leaderboard exactly for all \nmodels\ configurations.

Table~\ref{tab:ablation} shows the result. With zero near-clones retained, rank-1 retrieval is essentially solved (0.93--0.99 for the four leading systems; 0.532--0.993 across all \nmodels, 6 at or above 0.95) --- for the leading systems, topical retrieval is not the hard part. Retaining a \emph{single} near-clone, averaged over which one, drops the strongest system from 0.993 to 0.678; the relative one-clone drop is 27--49\% across all \nmodels. Density scales the effect but does not create it: the benchmark measures what happens \emph{given} a near-clone candidate in the pool, already at the lowest density.

\subsection{Canonical-vs-best-distractor similarity gap}
\label{sec:simgap}

Figure~\ref{fig:simgap} quantifies the mechanism. For each query, we compute (a) the cosine similarity between the query and the canonical, and (b) the maximum cosine over its paired mechanically mutated buggy variants. The right column of Figure~\ref{fig:simgap} plots the per-query difference (canonical $-$ best distractor); over the 938 queries with at least one mechanical-type paired distractor, the canonical is below at least one of those distractors in \textbf{66.8\%} of queries on Gemini Embedding~2 (median gap $-0.002$, mean $-0.002$) and \textbf{78.4\%} on Qwen3-Embedding-8B (median gap $-0.002$, mean $-0.002$). The mean per-query gap is small in absolute terms (the canonical and its best mutated variant are nearly indistinguishable to the embedding), but the sign is wrong on the majority of queries.

The two systems use different invocation recipes (Gemini uses a task-prefix instruction; Qwen uses an instruction-style prompt), yet both show the same qualitative pattern: overlapping cosine distributions for canonical vs.\ best paired distractor, with the per-query difference concentrated tightly around zero but biased toward the negative side. At least one of a query's near-clones typically embeds \emph{closer} to the NL query than the canonical, though an individual distractor does so in 44.3\% of triples.

\subsection{Deception is broadly distributed across bug types}
\label{sec:analysis-mechanism}

We define a (query, paired distractor, model) triple as \emph{deceived} when the distractor's similarity score to the query is at least as high as the canonical's (cosine over L$_2$-normalized embeddings for 22 of 23 models; raw dot product for \texttt{multi-qa-mpnet-base-dot-v1} per its model card). Across \nmodels\ dense configurations and the six mechanical mutation types of the locked prompt, 86{,}135 paired triples are evaluable, of which 44.3\% are deceived (Figure~\ref{fig:bugtype}). Per-bug-type rates span a narrow 39.3\%--48.0\% band: \texttt{remove\_edge\_case\_check} is the least deceptive (39.3\%) and \texttt{wrong\_comparison} the most (48.0\%); the other four cluster between 42.8\% and 47.0\%. No mechanical mutation is systematically near random; all six fool an embedding roughly two times in five. This argues that the failure mode is not bound to a single mutation pattern (e.g.\ off-by-one); it is a property of embedding near-identical code rather than of any particular bug.

Nor is the failure mode bound to micro-edits (Appendix~\ref{app:robustness}): deception shows no meaningful association with edit size (39.5--47.5\% across character-span buckets, 1 to 21+ characters; tie-aware Spearman $|\rho| \le 0.07$), and stacking 2--4 execution-verified, line-disjoint mutations into the same function leaves deception within 1.7 points of the single-mutation rate, with no drop statistically distinguishable from zero (matched 371-query comparison, four locally run embedders).

\subsection{Between-family gaps are detectable; within-family scaling is not}
\label{sec:pairwise}

\begin{table*}[!tb]
\centering
\footnotesize
\setlength{\tabcolsep}{3pt}
\renewcommand{\arraystretch}{0.9}
\begin{tabular}{l r r r r r r}
\toprule
 & \multicolumn{4}{c}{\texttt{exec@1} paired tests} & & \\
\cmidrule(lr){2-5}
Comparison ($A$ vs $B$) & $A$-only & $B$-only & $\Delta$\texttt{exec@1} & McNemar exact $p$ & $\Delta$\texttt{exec@3} & $\Delta$\texttt{exec@5}\\
\midrule
Gemini Emb.\ 2 vs Gemini Emb.\ 001 & 179 & 177 & $+0.002$ & 0.958 & $-0.066$ ($p$=$2.7\!\times\!10^{-5}$) & $+0.009$\\
Gemini Emb.\ 2 vs Mistral Embed & 224 & 123 & $+0.108$ & $6.5\!\times\!10^{-8}$ & $+0.028$ & $+0.060$\\
Gemini Emb.\ 2 vs Qwen3-8B & 228 & 118 & $+0.117$ & $3.4\!\times\!10^{-9}$ & $+0.039$ & $+0.018$\\
Qwen3-8B vs Qwen3-4B BF16 & 136 & 127 & $+0.010$ & 0.622 & $-0.002$ & $-0.004$\\
Qwen3-4B BF16 vs Qwen3-4B FP16 & 48 & 40 & $+0.009$ & 0.456 & $-0.005$ & $+0.001$\\
\bottomrule
\end{tabular}
\caption{Selected paired comparisons. ``$A$-only''/``$B$-only'' are McNemar discordant counts at $k=1$ out of \nqueries. The full pairwise table (every pair, all $k$, all three metrics) is released as \texttt{pairwise\_tests.json}.}
\label{tab:pairwise}
\end{table*}

Table~\ref{tab:pairwise} reports the headline paired tests. Gemini~2 and Gemini~001 are indistinguishable at rank~1 (paired McNemar exact $p=0.958$) but Gemini~001 is significantly stronger at \texttt{exec@3} ($p\!\approx\!2.7\!\times\!10^{-5}$) --- which is the operating regime where the choice between the two Google models matters. Gemini~2 is a clear winner over Mistral Embed and Qwen3-8B at \texttt{exec@1} (both $p<10^{-7}$); the absolute gap on Qwen3-8B is $+0.117$, i.e.\ Gemini~2 succeeds at rank~1 on $\sim$110 more queries than Qwen3-8B out of \nqueries. Within the Qwen3 family we detect no scaling: 8B is not significantly better than 4B-BF16 at \texttt{exec@1} ($p=0.622$), nor does 4B-BF16 outperform 4B-FP16 ($p=0.456$). We therefore report the family-level claim (Qwen3 leads the open-weight tier) rather than a specific scaling claim.

\subsection{Execution and identity ranking disagree}

Execution and identity-based retrieval are not interchangeable. Gemini~001 leads on nDCG@5 and on \texttt{exec@3}, but Gemini~2 has the highest \texttt{exec@1}; Codestral Embed 2505 reaches the highest \texttt{exec@5} among non-Gemini systems with lower nDCG@5. The median query is solved at rank~1 by only 3 of \ntotalmodels\ systems and 88 queries by no system (Appendix~\ref{app:difficulty}); the per-domain breakdown shows date-time as the hardest category (mean \texttt{exec@1}~$=$~0.091 across all 24 systems, vs.\ 0.107--0.151 elsewhere), bottom-three for every top-five model.

\section{Discussion and Conclusion}
\label{sec:conclusion}

With up to four near-clones planted per query, top-$k$ is strong (\texttt{exec@10}$\geq$0.98 for the four leading systems) but rank-1 is brittle: misses return a paired buggy variant 91.5--99.4\% of the time, and the canonical scores below at least one paired distractor in 67--78\% of queries on the leading systems. Current code embeddings function as candidate-recall components; both metric families matter. \bench\ thus measures the correctness-discrimination burden that candidate retrieval passes to downstream verification --- reranking, execution, or an LLM examining the forwarded candidates --- not the failure of complete coding-agent systems.

\clearpage
\section*{Limitations}
\label{sec:limitations}

\paragraph{Python only.} All canonicals, distractors, and tests are Python. Cross-language generalization remains open; the generation and validation pipeline is language-agnostic in design but each new language requires its own runner.

\paragraph{Mechanical, not human-written, distractors.} Our distractors are single-targeted-edit LLM mutations rather than natural human bugs (8.5\% of mechanical pairs span multiple lines, up to 15, predominantly edge-case-guard removals); correctness is relative to each entry's generated tests, which may exceed the query's stated domain. By design, mechanical mutations always differ from the canonical, the mutation family corresponds to classical mutation-testing operators \citep{jia2011mutation,just2014mutants}, and the locked prompt forbids LLM-style rewrites that often turn out to be functionally correct. Real-world bugs include subtler patterns and stylistic confounds this benchmark does not exhaust; Appendix~\ref{app:robustness} finds no detectable dependence of deception on edit size (23 configurations) or stacked-mutation count (four local embedders).

\paragraph{Closed-world corpus.} The retrieval corpus is exactly the \ncorpus\ snippets generated for the benchmark. We do not add open-domain code or distractors from other corpora. This isolates the functional-correctness signal but makes \texttt{exec@k} incomparable to open-domain code-search numbers. Every reported magnitude is likewise conditional on a near-clone candidate in the pool (Section~\ref{sec:ablation}); how often deployed corpora pose this choice is unmeasured --- doing so requires an execution oracle over open-domain code. Near-clones plausibly co-occur with their targets in agentic loops caching failed attempts, retrieval over several sampled implementations of a task, monorepos with copy-pasted diverged utilities, and commit-history indices holding pre-fix versions. Clone studies make this plausible: 70\% of GitHub code consists of file-level clones of earlier files \citep{lopes2017dejavu}, about half of clone groups contain inconsistently modified members \citep[52\%;][]{juergens2009clones}, replicated on three further industrial systems \citep{wagner2016clones}, and Python commit histories alone contain nearly a million deduplicated single-statement pre-fix variants of later-fixed code \citep{richter2022tssb,prenner2025bogus} --- though these establish plausibility, not frequency in retrieval results.

\paragraph{Execution runner is not a hardened sandbox.} The Python subprocess runner provides process isolation and a 5-second timeout, but does not constrain filesystem or network access. We do not recommend using it on untrusted code outside this benchmark.

\paragraph{Embedding-only first-stage retrieval.} \bench\ measures the first retrieval stage. The natural next question is how cross-encoder rerankers (or LLM rerankers) behave on these distractors; see Future work below.

\paragraph{Provider API drift.} Hosted-API embeddings are not permanently reproducible across provider model updates. We mitigate this by releasing the resulting embedding matrices and the execution cache; re-running scoring is fully deterministic from the saved \texttt{.npz} matrices, even if the live provider endpoint changes.

\paragraph{Scale.} \nqueries\ queries is small compared to topical retrieval corpora that contain millions of (query, function) pairs. The binding cost is per-query execution verification of every distractor against every test, and is intentional. The tasks are deliberately self-contained functions --- the level at which a uniform oracle and attributable single-edit failures are tractable; repository-scale code with cross-file context is beyond them. The pipeline scales with generation budget (Appendix~\ref{app:economics}).

\paragraph{Future work.} Three follow-ups are natural. First, cross-encoder and LLM-based rerankers operating on the embedding-retrieved top-$k$ may close the rank-1 gap without retraining the first-stage embedding models; the released oracle and frozen matrices make measuring that rank-1 lift deterministic. Second, porting the pipeline to JavaScript, Rust, or Go would test whether the near-clone deception we report is Python-specific or general across languages. Third, free-form LLM-generated buggy implementations, kept only when they fail the tests, would probe a broader, more naturalistic bug distribution; this complements rather than replaces mechanical mutation --- a free-form rewrite loses the single-attributable-edit control, and execution filtering is not optional (127 of 400 pilot buggy-rewrite requests were accidentally correct; Section~\ref{sec:generation}).

\section*{Acknowledgments}
We thank the three anonymous ARR reviewers and the area chair, whose requests and suggestions shaped this version: the pool-density, edit-size, and multi-mutation analyses were added during the review period at their prompting. This work received no institutional or grant funding; the first author thanks Manish Kapoor for funding this project and for his encouragement and thoughtful discussions throughout.

\bibliography{custom}

\appendix

\section{Generation Prompts and Example Entry}
\label{app:prompts}

Listings~\ref{lst:sysprompt}--\ref{lst:userprompt} reproduce the verbatim system prompt and user-prompt clauses used by the generation pipeline of Section~\ref{sec:generation}. Listing~\ref{lst:entry-example} shows a stylized excerpt of one validated entry (\texttt{next\_power\_of\_two\_bitwise}, query \texttt{q\_0134}, canonical \texttt{c\_0666}, distractors \texttt{c\_0667}--\texttt{c\_0670}); the full record (with all 9 tests and verbatim \texttt{bug\_description} strings) is in the released JSONL files.

\needspace{14\baselineskip}
\begin{lstlisting}[caption={System prompt for generation.},label={lst:sysprompt}]
You are an expert Python programmer
generating benchmark data for a code
retrieval evaluation system.
You must return ONLY valid JSON --
no markdown fences, no commentary,
no explanation. Just a JSON array.
\end{lstlisting}

\begin{lstlisting}[caption={User-prompt clauses (verbatim).},label={lst:userprompt}]
For each query, produce:
1. canonical: correct stdlib-only Python.
2. test_suite: 7-10 assert statements
   covering normal, edge, and
   distinguishing cases. STRICT FORMAT:
   each test MUST be a single
   `assert function_name(...) == ...`.
   No helpers, no setup, no comments.
3. distractors: 4 plausible-but-wrong
   functions, each a MECHANICAL MUTATION
   of the canonical (single targeted
   change). Must produce WRONG OUTPUT,
   not crash. Allowed bug types:
   off_by_one, wrong_operator,
   swap_arguments, remove_edge_case_check,
   wrong_comparison, off_by_one_boundary.
   DO NOT use "wrong_semantics" --
   writing an alternative correct
   implementation is not a valid
   distractor.
\end{lstlisting}

\begin{lstlisting}[caption={Stylized excerpt of one validated entry; full record in released JSONL files.},label={lst:entry-example}]
{
 "function_name": "next_power_of_two_bitwise",
 "query": "Return the smallest power of
  two >= a given positive integer n.
  Use bit manipulation, not math.log.",
 "canonical":
   "def next_power_of_two_bitwise(n):\n
    if n < 2:\n        return 1\n
    n -= 1\n    shift = 1\n
    while (n >> shift) > 0:\n
        n |= n >> shift\n
        shift <<= 1\n    return n + 1",
 "test_suite": [
   "assert next_power_of_two_bitwise(0)==1",
   "assert next_power_of_two_bitwise(1)==1",
   ...
 ],
 "distractors": [
   {"code": "...",
    "bug_type": "remove_edge_case_check",
    "bug_description":
       "Deleted the initial `if n < 2:`."},
   {"code": "...",
    "bug_type": "wrong_comparison",
    "bug_description":
       "Changed `n < 2` to `n <= 2`."},
   {"code": "...",
    "bug_type": "off_by_one_boundary",
    "bug_description":
       "Changed `n -= 1` to `n -= 2`."},
   {"code": "...",
    "bug_type": "wrong_operator",
    "bug_description":
       "Final return `n + 1` -> `n - 1`."}
 ]
}
\end{lstlisting}

\section{Generation Economics}
\label{app:economics}

We aggregate the per-entry \texttt{batch\_usage} metadata of the 926 released entries with available token counts. GPT-5.4 with \texttt{reasoning\_effort=high} consumed, in total, \textbf{915{,}178 prompt tokens and 7{,}083{,}062 completion tokens} across these 926 API calls (5{,}871{,}712 of which are hidden reasoning tokens, reported by OpenAI as a subset of \texttt{completion\_tokens}; mean per entry: 988 prompt, 7{,}649 completion of which 6{,}341 reasoning; median wall-clock 99~s for real-time-endpoint entries; batch-endpoint entries do not record per-request latency). At GPT-5.4 list pricing of \$2.50 per 1M input tokens and \$15.00 per 1M output tokens (output equals \texttt{completion\_tokens}, which already includes reasoning per OpenAI's billing convention; pricing and billing conventions per the provider documentation \citep{openai_gpt54}, as of dataset construction), the generation cost at native real-time list pricing is \textbf{\$108.53}. Approximately 57\% of these API calls were routed through the OpenAI Batch API, which carries a 50\% discount on both rates and a 24-hour SLA; we report the list-price figure as the headline cost since it is a lower bound on the budget required to reproduce the dataset without depending on batch-endpoint availability. With either prompt fix from Section~\ref{sec:generation} missing, end-to-end validation rates fall to 62--70\% even with high-effort reasoning, so the cost reported here would scale roughly inversely with the validation rate.

\section{Per-Query Difficulty and Per-Domain exec@1}
\label{app:difficulty}

\begin{center}
\includegraphics[width=\columnwidth]{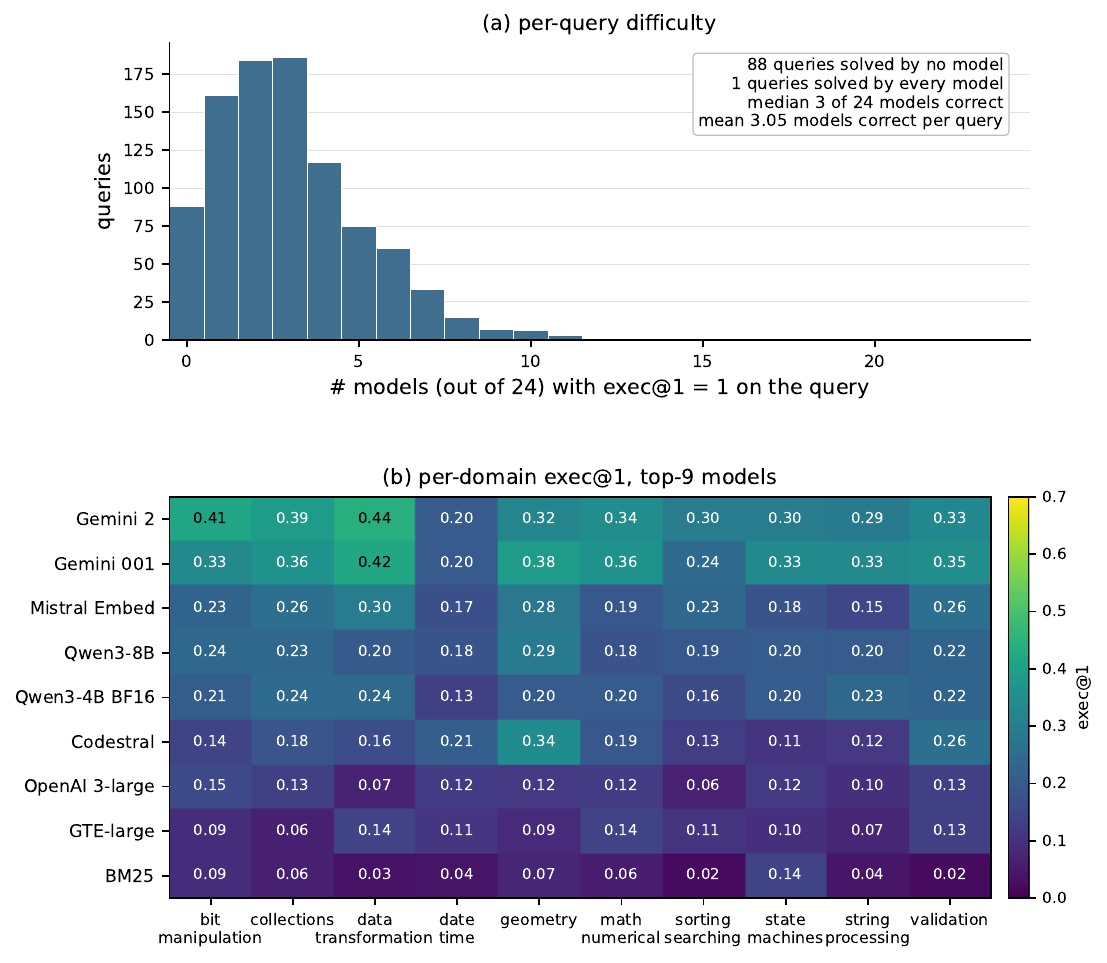}
\captionof{figure}{(a) Distribution of how many of the \ntotalmodels\ evaluated systems achieve \texttt{exec@1}$=$1 on each query. The median query is solved by exactly 3 of 24 systems; 88 queries are solved by no system, 1 by every system. (b) Per-domain \texttt{exec@1} for the top-9 systems.}
\label{fig:difficulty}
\end{center}

\begin{table*}[!t]
\centering
\footnotesize
\renewcommand{\arraystretch}{1.05}
\setlength{\tabcolsep}{3pt}
\begin{tabular}{@{}>{\raggedright\arraybackslash}p{3.0cm} l l c l l >{\raggedright\arraybackslash}p{3.6cm}@{}}
\toprule
Model & Provider & dim & sim & batch & dtype & Query / passage conditioning\\
\midrule
Gemini Emb.\ 001 & Google nat. & 3072 & cos & 32 & -- & task type \texttt{CODE\_RETRIEVAL\_QUERY} / \texttt{RETRIEVAL\_DOCUMENT}\\
Gemini Emb.\ 2 & Google nat. & 3072 & cos & 32 & -- & ``task: code retrieval $|$ \mbox{query: ''} / ``title: none $|$ \mbox{text: ''}\\
Mistral Embed & Mistral nat. & 1024 & cos & 64 & -- & raw text\\
Codestral Embed 2505 & Mistral nat. & 1536 & cos & 64 & -- & raw text\\
OpenAI 3-L / 3-S / ada-002 & OpenAI nat. & 3072/1536/1536 & cos & 128 & -- & raw text\\
Qwen3 4B / 4B BF16 / 8B & Local HF & 2560/2560/4096 & cos & 8/8/4 & fp16/bf16/bf16 & ``Instruct: $\langle$task$\rangle$\textbackslash nQuery: '' on queries\\
BGE-base / -large v1.5 & Local HF & 768/1024 & cos & 64/32 & -- & ``Represent this sentence for searching relevant \mbox{passages: ''}\\
BGE-M3 (dense only) & Local HF & 1024 & cos & 32 & -- & native FlagEmbedding dense path\\
E5-base/-large-v2, mE5-large & Local HF & 768/1024/1024 & cos & 64/32/32 & -- & ``query: '' / ``passage: ''\\
GTE-base / -large & Local HF & 768/1024 & cos & 64/32 & -- & raw text\\
S-T MiniLM-L6 / MiniLM-L12 / MPNet / Para-MiniLM & Local HF & 384/384/768/384 & cos & 128/128/64/128 & -- & native sentence-transformers\\
Multi-QA MPNet & Local HF & 768 & dot & 64 & -- & native; no L2 normalization\\
BM25 & lexical & -- & -- & -- & -- & \texttt{k1=1.5}, \texttt{b=0.75}\\
\bottomrule
\end{tabular}
\caption{Provider-native and local-best invocation settings used in \bench.}
\label{tab:invocation}
\end{table*}

\section{Released Artifacts}
\label{app:artifacts}

The full release lives at a Hugging Face dataset: \url{https://huggingface.co/datasets/AaryanK/ExecRetrieval}. All files are released under CC~BY~4.0. The bundle contains:
\begin{enumerate}
\item the corpus (\texttt{corpus.jsonl}, \ncorpus\ rows) and the query file (\texttt{queries.jsonl}, \nqueries\ rows), with deterministic IDs and per-entry \texttt{metadata} (generator model id, endpoint, generation timestamp) plus \texttt{batch\_usage} (prompt/completion/reasoning tokens and latency) for the 926 instrumented entries of Appendix~\ref{app:economics};
\item the per-distractor metadata file with the \texttt{bug\_description} explanation field for each of the \ndistractors\ paired distractors;
\item the per-pair execution cache (46{,}458 outcome rows keyed by \texttt{(code\_sha256, tests\_sha256)});
\item the 23 saved \texttt{.npz} embedding matrices (one per dense embedding configuration);
\item the full leaderboard table (Table~\ref{tab:full-leaderboard}) and the full raw pairwise-tests JSON (every model-pair comparison for $k\in\{1,3,5,10\}$ on \texttt{exec@k}, \texttt{execution\_precision@k}, and nDCG@$k$);
\item the runtime \texttt{pip freeze};
\item a SHA-256 manifest covering all of the above;
\item the evaluation harness (\texttt{eval/}: subprocess executor, scoring engine, BM25 baseline, embedder scripts, pairwise tests), including \texttt{eval/reproduce\_leaderboard.sh}, a single command that recomputes the entire leaderboard offline from the frozen matrices and diffs it against the released results, and the construction pipeline (\texttt{generation/}: 954-entry registry, LLM generation scripts for both real-time and OpenAI Batch endpoints, five-gate validator, corpus builder) so that both the scoring step of Section~\ref{sec:protocol} and the construction procedure of Sections~\ref{sec:generation}--\ref{sec:validation} are independently re-runnable;
\item the analysis scripts and outputs for Section~\ref{sec:ablation} and Appendix~\ref{app:robustness} (\texttt{rebuttal/}: the pool-density ablation, edit-size, and mutation-stacking analyses).
\end{enumerate}
Metrics recompute deterministically from the frozen matrices without provider API calls.

\section{Per-Model Invocation Table}
\label{app:invocation}

Table~\ref{tab:invocation} reproduces the operationally critical invocation settings for every evaluated system. Provider documentation we used to fix invocation conventions: Google Gemini Embedding (\url{https://ai.google.dev/gemini-api/docs/embeddings} \citep{google_gemini_embeddings_docs}); OpenAI Embeddings (\url{https://platform.openai.com/docs/guides/embeddings} \citep{openai_embeddings_docs}); Mistral Embeddings (\url{https://docs.mistral.ai/api/endpoint/embeddings} \citep{mistral_embeddings_docs}); Qwen3 Embedding model cards on Hugging Face (\texttt{Qwen/Qwen3-Embedding-\{4B,8B\}}); BAAI BGE model cards (\texttt{BAAI/bge-\{base,large\}-en-v1.5}, \texttt{BAAI/bge-m3}); E5 model cards (\texttt{intfloat/e5-\{base,large\}-v2}, \texttt{intfloat/multilingual-e5-large}); GTE model cards (\texttt{thenlper/gte-\{base,large\}}); Sentence-Transformers model cards on Hugging Face for the remaining models.

\section{Full Per-Model Leaderboard}
\label{app:full-leaderboard}

Table~\ref{tab:full-leaderboard} reports \texttt{exec@k} for every system and $k\in\{1,3,5,10\}$ so that every figure-level claim is verifiable from the paper alone.

\input{tables/full_leaderboard}

\section{Robustness Analyses: Edit Size and Stacked Mutations}
\label{app:robustness}

Two further analyses test whether the deception measurement of Section~\ref{sec:analysis-mechanism} is an artifact of micro-edits. Both are derived from the released artifacts; the analysis scripts ship with the release.

\subsection{Deception vs.\ edit size}
\label{app:editsize}

For each of the 3{,}745 released mechanical (canonical, distractor) pairs (the six locked-prompt mutation types; the 10 pre-lock legacy distractors are excluded), we measure the number of characters spanned by the non-equal regions of a character-level diff. Edit sizes range from 1 to 540 characters; the median mutation changes a \emph{single character}, so we report fixed buckets rather than quantiles. Aggregated over the \nmodels\ dense configurations (86{,}135 triples; aggregate deception 44.3\%, matching Section~\ref{sec:analysis-mechanism}):

\begin{table}[htbp]
\centering
\small
\begin{tabular}{lrr}
\toprule
Edit size & pairs & deception\\
\midrule
1 character & 1{,}894 & 43.4\%\\
2--5 characters & 977 & 46.4\%\\
6--20 characters & 513 & 47.5\%\\
21+ characters & 361 & 39.5\%\\
\bottomrule
\end{tabular}
\caption{Deception by edit size over the 3{,}745 mechanical pairs $\times$ 23 dense configurations.}
\label{tab:editsize}
\end{table}

Deception is flat across edit sizes. The tie-aware Spearman correlation between edit size and per-pair deception is $|\rho| \le 0.07$ (relative edit size: $\rho = -0.069$; absolute: $\rho = +0.035$) --- statistically significant at $n = 3{,}745$ but negligible in magnitude. The dip in the largest bucket is consistent with its composition: multi-line edits (8.5\% of mechanical pairs, up to 15 lines) are predominantly \texttt{remove\_edge\_case\_check} mutations, the least deceptive type overall (39.3\%, Figure~\ref{fig:bugtype}).

\subsection{Deception vs.\ number of stacked mutations}
\label{app:stacking}

Distractors whose mutations occupy disjoint line regions of the same canonical can be composed. We take every mechanical distractor whose diff against its canonical is a single contiguous line block, stack 2--4 blocks with pairwise-disjoint line ranges (deterministic per-query subsampling), and execute every composed variant against its query's full test suite in the same isolated-subprocess runner as the release. Compositions that accidentally pass all tests are discarded (13 of 4{,}646), leaving 4{,}633 execution-verified multi-mutation variants covering 915 of \nqueries\ queries.

We embed the composed variants with four of the locally run leaderboard embedders (\texttt{all-mpnet-base-v2}, \texttt{bge-base-en-v1.5}, \texttt{e5-base-v2}, \texttt{gte-large}) under the release's exact passage-side conventions --- verified by re-embedding released corpus rows and requiring cosine $\ge 0.999$ against the frozen matrices (observed $\ge 0.999999$) --- while query and canonical similarities come from the frozen matrices themselves. The remaining systems are not re-embedded; the analysis was scoped to these four local embedders. Queries admitting four disjoint edits are not a random subset (their single-mutation deception runs 1.5--1.6 points above the corpus-wide rate pooled over these four models), so we compare on the 371 queries with validated variants at \emph{every} mutation count, aggregating per query first and then averaging over the four models (Table~\ref{tab:stacking}):

\begin{table}[htbp]
\centering
\small
\begin{tabular}{lcc}
\toprule
Mutations & deception & $\Delta$ vs.\ 1 (95\% CI)\\
\midrule
1 & 49.0\% & ---\\
2 & 48.2\% & $-0.7$ [$-2.1$, $+0.6$]\\
3 & 47.3\% & $-1.6$ [$-3.5$, $+0.3$]\\
4 & 47.6\% & $-1.4$ [$-3.7$, $+0.9$]\\
\bottomrule
\end{tabular}
\caption{Deception vs.\ number of stacked mutations on the matched 371-query subset (per-query means averaged over four local embedders); $\Delta$ in percentage points with query-level bootstrap 95\% CIs. $\Delta$ is computed before rounding.}
\label{tab:stacking}
\end{table}

Deception stays within 1.7 points of the single-mutation baseline, and no drop is statistically distinguishable from zero (query-level bootstrap, 5{,}000 replicates; $\Delta$ in percentage points); even the most pessimistic interval edge leaves deception near 45\%. The 49.0\% baseline here differs from Section~\ref{sec:analysis-mechanism}'s 44.3\% for two reasons, both compositional: the matched 371-query subset is biased upward (49.0\% vs.\ 47.4\% for the same four models unrestricted), and the four local embedders differ from the \nmodels-model pool (47.4\% vs.\ 44.3\%). Stacking up to four verified bugs into one function does not restore detectability for these embedders.

\end{document}

%% file: tables/full_leaderboard.tex
\begin{table*}[!t]
\centering
\footnotesize
\setlength{\tabcolsep}{4pt}
\begin{tabular}{l rrrr rrrr rrrr}
\toprule
 & \multicolumn{4}{c}{\texttt{exec@k}} & \multicolumn{4}{c}{\texttt{execution\_precision@k}} & \multicolumn{4}{c}{nDCG@k}\\
\cmidrule(lr){2-5}\cmidrule(lr){6-9}\cmidrule(lr){10-13}
Model & $k$=1 & $k$=3 & $k$=5 & $k$=10 & $k$=1 & $k$=3 & $k$=5 & $k$=10 & $k$=1 & $k$=3 & $k$=5 & $k$=10\\
\midrule
Gemini Emb.\ 2 & 0.331 & 0.823 & 0.997 & 1.000 & 0.331 & 0.274 & 0.199 & 0.100 & 0.331 & 0.614 & 0.687 & 0.688 \\
Gemini Emb.\ 001 & 0.329 & 0.889 & 0.988 & 1.000 & 0.329 & 0.296 & 0.198 & 0.100 & 0.329 & 0.656 & 0.698 & 0.702 \\
Mistral Embed & 0.224 & 0.795 & 0.937 & 0.984 & 0.224 & 0.265 & 0.187 & 0.098 & 0.224 & 0.552 & 0.612 & 0.627 \\
Qwen3-8B & 0.214 & 0.784 & 0.979 & 0.997 & 0.214 & 0.261 & 0.196 & 0.100 & 0.214 & 0.539 & 0.621 & 0.627 \\
Qwen3-4B BF16 & 0.204 & 0.786 & 0.983 & 0.998 & 0.204 & 0.262 & 0.197 & 0.100 & 0.204 & 0.535 & 0.618 & 0.623 \\
Qwen3-4B FP16 & 0.196 & 0.791 & 0.982 & 0.998 & 0.196 & 0.264 & 0.196 & 0.100 & 0.196 & 0.535 & 0.616 & 0.621 \\
Codestral Embed 2505 & 0.185 & 0.781 & 0.988 & 1.000 & 0.185 & 0.260 & 0.198 & 0.100 & 0.185 & 0.521 & 0.609 & 0.613 \\
OpenAI 3-large & 0.113 & 0.701 & 0.939 & 0.985 & 0.113 & 0.234 & 0.188 & 0.099 & 0.113 & 0.443 & 0.544 & 0.558 \\
OpenAI ada-002 & 0.104 & 0.662 & 0.904 & 0.965 & 0.104 & 0.221 & 0.181 & 0.097 & 0.104 & 0.420 & 0.521 & 0.540 \\
GTE-large & 0.104 & 0.661 & 0.897 & 0.959 & 0.104 & 0.220 & 0.179 & 0.096 & 0.104 & 0.415 & 0.514 & 0.534 \\
MiniLM-L12 & 0.086 & 0.564 & 0.832 & 0.909 & 0.086 & 0.188 & 0.166 & 0.091 & 0.086 & 0.354 & 0.466 & 0.491 \\
mE5-large & 0.085 & 0.622 & 0.840 & 0.932 & 0.085 & 0.207 & 0.168 & 0.093 & 0.085 & 0.385 & 0.476 & 0.506 \\
BGE-M3 (dense) & 0.085 & 0.571 & 0.811 & 0.888 & 0.085 & 0.190 & 0.162 & 0.089 & 0.085 & 0.356 & 0.457 & 0.481 \\
OpenAI 3-small & 0.083 & 0.624 & 0.863 & 0.941 & 0.083 & 0.208 & 0.172 & 0.094 & 0.083 & 0.383 & 0.483 & 0.508 \\
BGE-base & 0.083 & 0.626 & 0.862 & 0.943 & 0.083 & 0.209 & 0.172 & 0.094 & 0.083 & 0.387 & 0.486 & 0.512 \\
BGE-large & 0.078 & 0.630 & 0.859 & 0.941 & 0.078 & 0.210 & 0.172 & 0.094 & 0.078 & 0.385 & 0.481 & 0.507 \\
E5-large & 0.077 & 0.636 & 0.858 & 0.950 & 0.077 & 0.212 & 0.172 & 0.095 & 0.077 & 0.389 & 0.483 & 0.512 \\
MPNet-base & 0.076 & 0.603 & 0.851 & 0.939 & 0.076 & 0.201 & 0.170 & 0.094 & 0.076 & 0.370 & 0.474 & 0.502 \\
Multi-QA MPNet & 0.073 & 0.535 & 0.800 & 0.897 & 0.073 & 0.178 & 0.160 & 0.090 & 0.073 & 0.330 & 0.441 & 0.472 \\
GTE-base & 0.071 & 0.627 & 0.887 & 0.952 & 0.071 & 0.209 & 0.177 & 0.095 & 0.071 & 0.382 & 0.491 & 0.512 \\
MiniLM-L6 & 0.069 & 0.571 & 0.800 & 0.892 & 0.069 & 0.190 & 0.160 & 0.089 & 0.069 & 0.348 & 0.444 & 0.474 \\
E5-base & 0.069 & 0.609 & 0.863 & 0.935 & 0.069 & 0.203 & 0.172 & 0.093 & 0.069 & 0.371 & 0.477 & 0.500 \\
BM25 (lexical) & 0.058 & 0.234 & 0.328 & 0.422 & 0.058 & 0.078 & 0.066 & 0.042 & 0.058 & 0.159 & 0.198 & 0.228 \\
Para-MiniLM & 0.050 & 0.383 & 0.556 & 0.671 & 0.050 & 0.128 & 0.111 & 0.067 & 0.050 & 0.235 & 0.308 & 0.344 \\
\bottomrule
\end{tabular}
\caption{Full per-model leaderboard. exec@$k$ is the fraction of queries with at least one passing snippet in the top $k$; execution\_precision@$k$ is the fraction of top-$k$ snippets that pass; nDCG@$k$ scores rank quality against canonical-ID matches. Sorted by exec@1 descending. At $k=1$ the three metrics coincide.}
\label{tab:full-leaderboard}
\end{table*}